\documentclass[fleqn,usenatbib]{mnras}

\usepackage{newtxtext,newtxmath}
\usepackage{graphicx}
\usepackage{amsmath}	
\usepackage{gensymb}
\usepackage{subcaption}

\usepackage[T1]{fontenc}

\DeclareRobustCommand{\VAN}[3]{#2}
\let\VANthebibliography\thebibliography
\def\thebibliography{\DeclareRobustCommand{\VAN}[3]{##3}\VANthebibliography}

\usepackage{graphicx}	
\usepackage{amsmath}	
\usepackage{tikz,xcolor,hyperref}

\definecolor{lime}{HTML}{A6CE39}
\DeclareRobustCommand{\orcidicon}{
	\begin{tikzpicture}
	\draw[lime, fill=lime] (0,0) 
	circle [radius=0.16] 
	node[white] {{\fontfamily{qag}\selectfont \tiny ID}};
	\draw[white, fill=white] (-0.0625,0.095) 
	circle [radius=0.007];
	\end{tikzpicture}
	\hspace{-2mm}
}

\foreach \x in {A, ..., Z}{\expandafter\xdef\csname orcid\x\endcsname{\noexpand\href{https://orcid.org/\csname orcidauthor\x\endcsname}
			{\noexpand\orcidicon}}
}

\title[A new relic in A2108]{A new enigmatic radio relic in the low mass cluster Abell 2108}

\author[Chatterjee et al.]{
Swarna Chatterjee,$^{1}$\thanks{E-mail: chatterjee.swarna1605@gmail.com}\orcidA{}, Majidul Rahaman,$^{2}$\orcidB{}, Abhirup Datta,$^{1}$\orcidC{}, Ruta Kale$^{3}$\orcidD{}, and Surajit Paul$^{4}$\thanks{E-mail: surajit.paul@manipal.edu}\orcidE{}
\\
$^{1}$Department of Astronomy, Astrophysics and Space Engineering, Indian Institute of Technology Indore, Indore, MP, India\\
$^{2}$Institute of Astronomy, National Tsing Hua University, Hsinchu 300044, Taiwan, R.O.C.\\
$^{3}$National Centre for Radio Astrophysics (NCRA), Tata Institute of Fundamental Research (TIFR), Pune 411007,
India\\
$^{4}$Manipal Centre for Natural Sciences, Centre of Excellence, Manipal Academy of Higher Education, Manipal, Karnataka 576104, India
}

\date{Accepted XXX. Received YYY; in original form ZZZ}

\pubyear{2023}

\begin{document}
\label{firstpage}
\pagerange{\pageref{firstpage}--\pageref{lastpage}}
\maketitle

\begin{abstract}
We report the discovery of a radio relic in the northeastern periphery of the cluster Abell~2108 (A2108). A2108 is a part of the uGMRT LOw-MAss Galaxy Cluster Survey (GLOMACS), where our main aim is to search for diffuse radio emission signatures in very sparsely explored low-mass galaxy clusters using uGMRT band-3 (central frequency 400 MHz). We used our uGMRT band-3 data along with the existing archival band-3 uGMRT data to improve image sensitivity. Along with the previously reported southwestern relic, the discovery of the new relic makes A2108 one of the few low-mass clusters hosting double relics. The new relic spans over a region of 610~kpc $\times$ 310~kpc and, interestingly, differs considerably in size and morphology from the other relic. With XMM-Newton science archive data, we also report the tentative detection of a mildly supersonic shock of Mach number $\mathcal{M}_\mathrm{SB}=1.42$ and  $\mathcal{M}_\mathrm{T} = 1.43$ from the surface brightness and temperature discontinuities, respectively near this newly found relic. Both the relics in A2108 are found to be significantly under-luminous compared to other double relic systems in the mass-luminosity plane. Though mild supersonic shocks resulting from an off-axis merger could have influenced their origin, we hypothesize that further local environments have played a crucial role in shaping their morphology.
\end{abstract}




\section{Introduction}

\begin{figure*}
\begin{center}
    \includegraphics[width=0.49\linewidth]{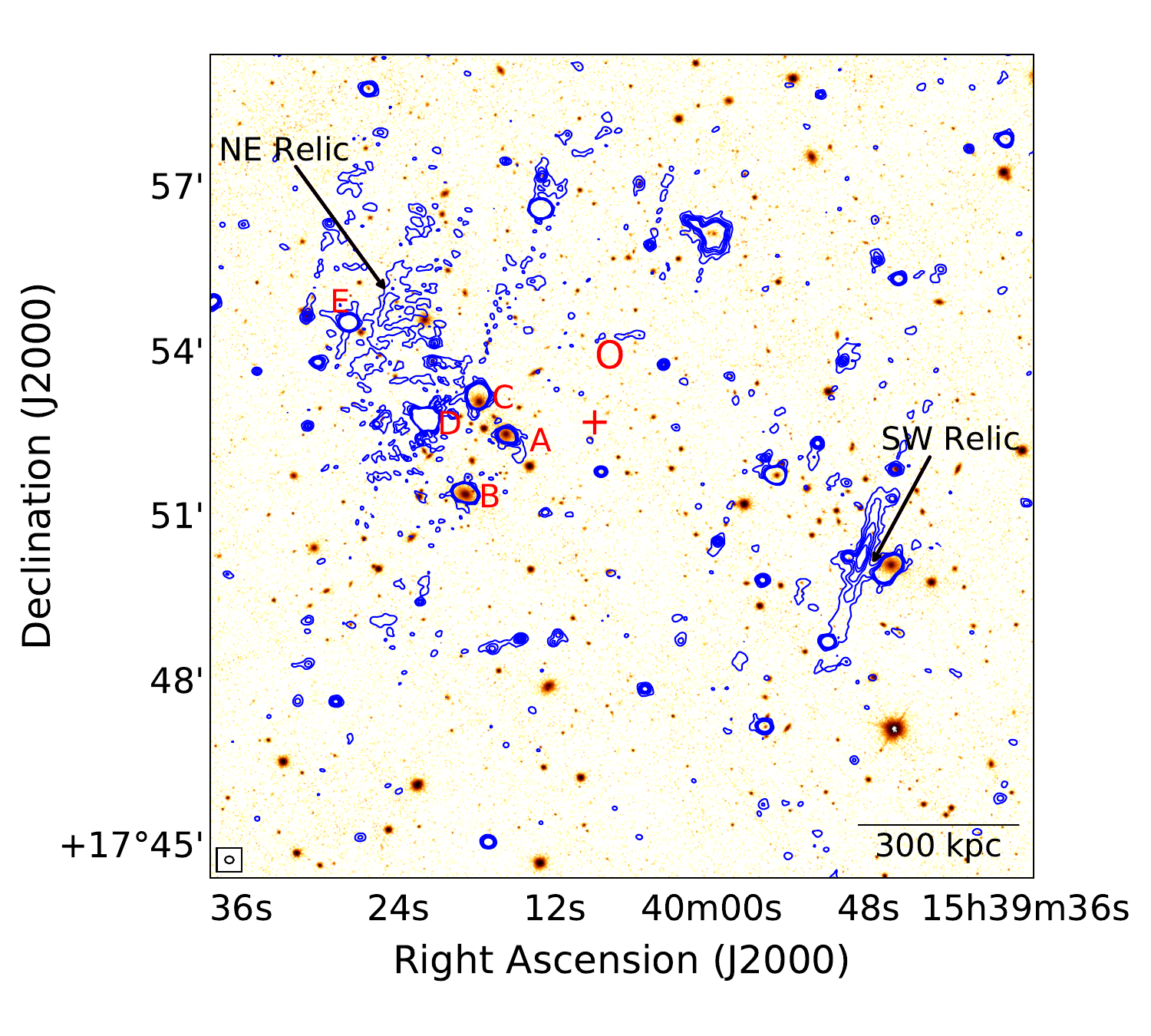}
    \includegraphics[width=0.49\linewidth]{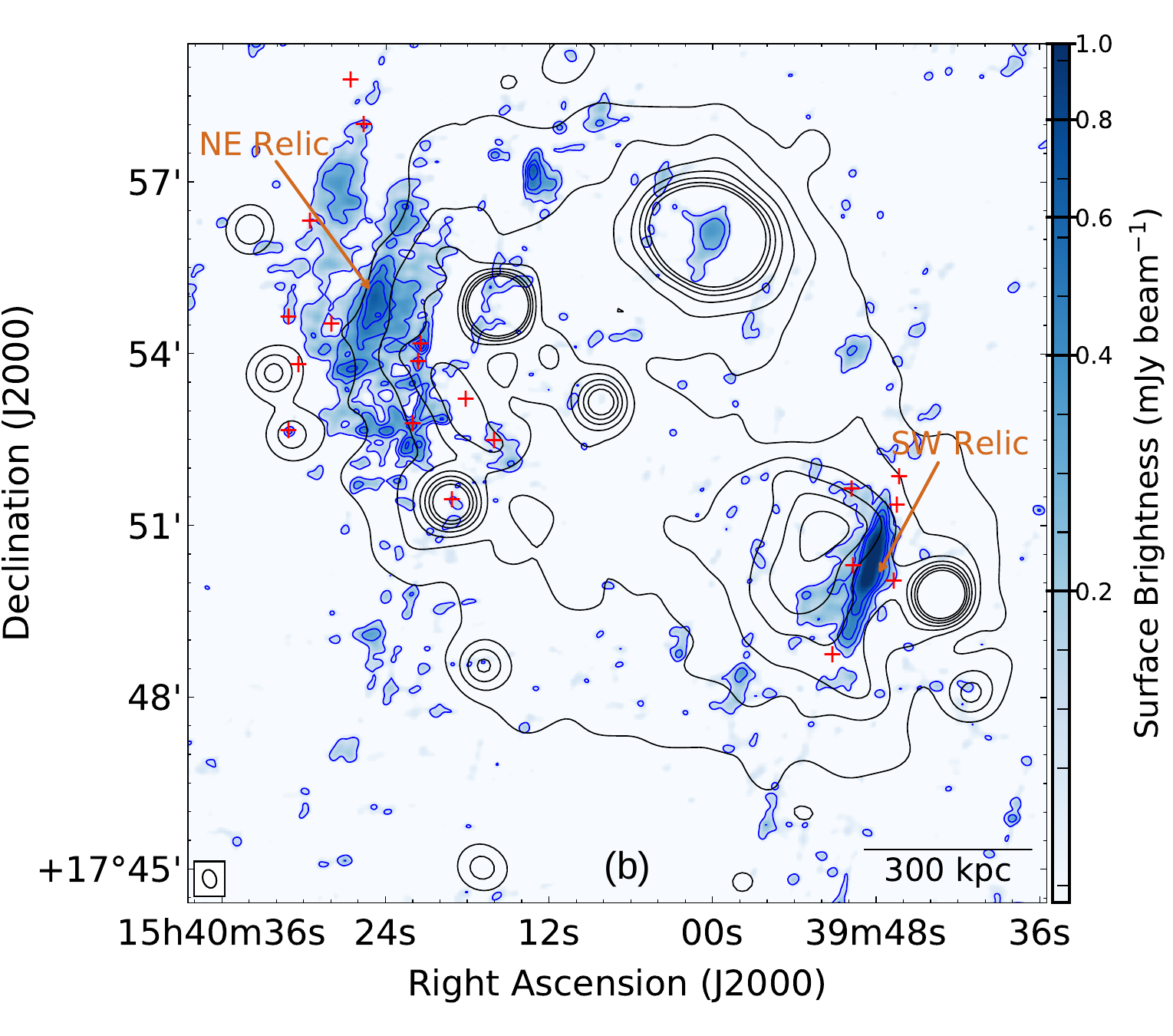}
\caption{(a): uGMRT 400~MHz blue contours with restoring beam $9.86\arcsec \times8.10\arcsec, $ PA 87.28$\degree$ overlaid on Pan-STARRS `$r$' band optical image. The contours are placed at (3,6,9,12,15)$\times 35~\mu$Jy/beam. The X-ray and SZ centres of the cluster are marked in red as + and O, respectively, and radio galaxies near NE relic at cluster redshift are marked from A-E. (b): XMM X-ray contours in black and the uGMRT 400~MHz point source subtracted radio contours in blue overlaid on the uGMRT 400~MHz point source subtracted colour image with restoring beam $19.59\arcsec \times14.13\arcsec, $ P.A 17.24$\degree$. The radio contours are placed at (3,6,9,12,15)$\times 42~\mu$Jy/beam. Subtracted compact sources near the diffuse emissions are highlighted with + in red.}
\label{fig: A2108_radio_xray}
\end{center}
\end{figure*}

Various cluster-scale diffuse radio emission has been reported in a number of galaxy clusters, elucidating the presence of magnetic fields and GeV electrons in the intracluster medium  (ICM; \citealt{Feretti_2012A&ARv..20...54F,vanweeren_2019SSRv..215...16V,paul_2023JApA}). The most spectacular among them are the Mpc scale radio relics also known as radio gischts, usually found either in a pair or as isolated ones at the outskirts or peripheral region of merging clusters. They are non-thermal synchrotron emission structures and are understood to be the by-product of Mpc scale merger shocks produced during cluster mergers \citep{Ensslin_1998A&A...332..395E, Roettiger1999ApJ}. In principle, the huge cluster merger energy ($E_\mathrm{merge}\sim10^{64}$~ergs) dissipates through the acceleration of particles, adiabatic heating, and possibly by amplifying magnetic fields in the dilute cluster medium \citep{Brunetti_2014IJMPD..2330007B}.

Radio relics are generally formed perpendicular to the merging axis of the clusters and are often seen to trace cluster merger shocks. These radio relics are therefore conceived to form by in-situ particle acceleration by the passage of merger shocks. Relics are found to be polarised ($>$30\%) at GHz frequencies with the magnetic field vectors aligned with the shock plane \citep{DiGennaro2021ApJ, Rajpurohit_2022A&A}
The radio spectral index distribution\footnote{We define the spectral index, $\alpha$, via $S_\nu \propto \nu^\alpha$ for flux density $S_\nu$ at frequency $\nu$} of relics shows a steepening from the inner to the outer edge. This phenomenon is well explained by the Diffusive Shock Acceleration (DSA) model, where the merger shock accelerates the particles at the shock fronts, producing a flatter spectrum near the shock position. \citep{Drury_1984AdSpR, Hoeft_2007MNRAS.375...77H, Kang_2013ApJ}.
However, this mechanism is found to be inefficient with low Mach number shocks; the Mach numbers are often estimated from associated X-ray density and temperature jumps \citep{vanWeeren_2017NatAs, Botteon_2020A&A}. Recent works, both from simulations and observations, have shown that the particle re-acceleration of pre-existing mildly relativistic electrons is a better approach for some of the observed radio relics \citep{Caprioli_2014ApJ...783...91C, Botteon_2016MNRAS.460L..84B, Gennaro_2018, Fernandez_2021MNRAS.500..795D}.

Due to both theoretical and instrumental limitations, the previous studies have been mostly biased toward the high mass clusters ($M_{500}\geq 5 \times 10^{15}~\rm{M}_{\odot}$; see \citealt{gasperin_2014MNRAS, vanweeren_2019SSRv..215...16V} for a review). With the recently increased sensitivity of radio telescopes, a significant number of relics in lower mass clusters are being discovered  (see \citealt{Botteon_2022A&A, paul_2023JApA} for a review). It is important to have information on radio relics from clusters in a wide mass range as the environment and the effectiveness of various particle acceleration engines may largely differ. 

\textbf{Abell 2108} (hereafter A2108) is a radio-relic hosting cluster in our uGMRT LOw-MAss Galaxy Cluster Survey (GLOMACS) sample list. A brief of the available information regarding the cluster is presented in Table~\ref{tab: Table1}. 
 
Notably, this is an X-ray under-luminous 
cluster in spite of having high ICM velocity dispersion \citep{Jensen_2012MNRAS}. 
The cluster hosts 3 brightest cluster galaxies (BCG) in the northern region itself \citep{Crawford_1999MNRAS,hogan_2015}. Recently, a steep spectrum radio relic ($\alpha = -2$) was discovered at the southwest of this cluster by \citet{Schellenberger_2022ApJ}. They also detected a shock $\sim200$ kpc away in the southwestern direction from the radio relic. In this paper, we report important new findings from the deeper analysis of uGMRT band-3 and XMM-Newton X-ray archival observation of A2108.

\begin{table}
	\caption{Properties of A2108 \citep{Jensen_2012MNRAS,Planck_Collaboration_2014A&A}}
	\label{tab: Table1}
	\begin{tabular}{lr} 
		\hline
		$\mathrm{RA_{J2000}}$ & 15 h 40 m 07.96 s \\
		$\mathrm{DEC_{J2000}}$ &  $+17^{\circ} 53\arcmin 55.2\arcsec$\\
		Mass ($\mathrm{M_{500}}$) & $3.01^ {+0.38}_{-0.40}\times10^{14}~\mathrm{M_\odot}$ \\
		Redshift (z) & 0.0916\\
        Dispersion velocity ($\mathrm{\sigma_z}$) & 750 km.s$^{-1}$\\
        $\mathrm{n_H}$ & $2.38 \times 10^{20}$ cm$^{-2}$ \\ 
		Luminosity ($\mathrm{L_{x (0.1-2.4)~keV}}$) & $1.97 \times 10^{44}~\mathrm{erg.s^{-1}}$\\
		\hline
	\end{tabular}
    
\end{table}

We assume a $\Lambda$CDM cosmology with $\Omega_m = 0.3$, $\Omega_{\Lambda}=0.7$ and $H_0=70$~km s$^{-1}$~Mpc$^{-1}$. At the cluster redshift of $z = 0.0916$, {1\arcsec} corresponds to 1.705781~kpc.
\section{Radio Observation}\label{obs}

\begin{table*}
	\centering
	\caption{Best fit parameters of broken power law model (using PROFFIT) are listed below.}
	\label{tab:Table2}
	\resizebox{0.9\textwidth}{!}{
	\begin{tabular}{lcccccc} 
		\hline
		\hline
		$\alpha1$ & $\alpha2$ & $r_\mathrm{sh}$(arcmin) & norm & Jump ($C$) & $\chi^2/D.o.f$ & $\mathcal{M}_\mathrm{SB}$\\
		\hline
		-1.14 $\pm$ 0.28 & 2.50 $\pm$ 0.80 & 5.32 $\pm$ 0.05 & $-2.85 \pm  0.05  e-05$ & 1.61 $\pm$ 0.27 & 1.50 & $1.42^{+0.14}_{-0.14}$ \\
		\hline
	\end{tabular}
	}
\end{table*}

\begin{figure*}
    \includegraphics[width=\columnwidth]{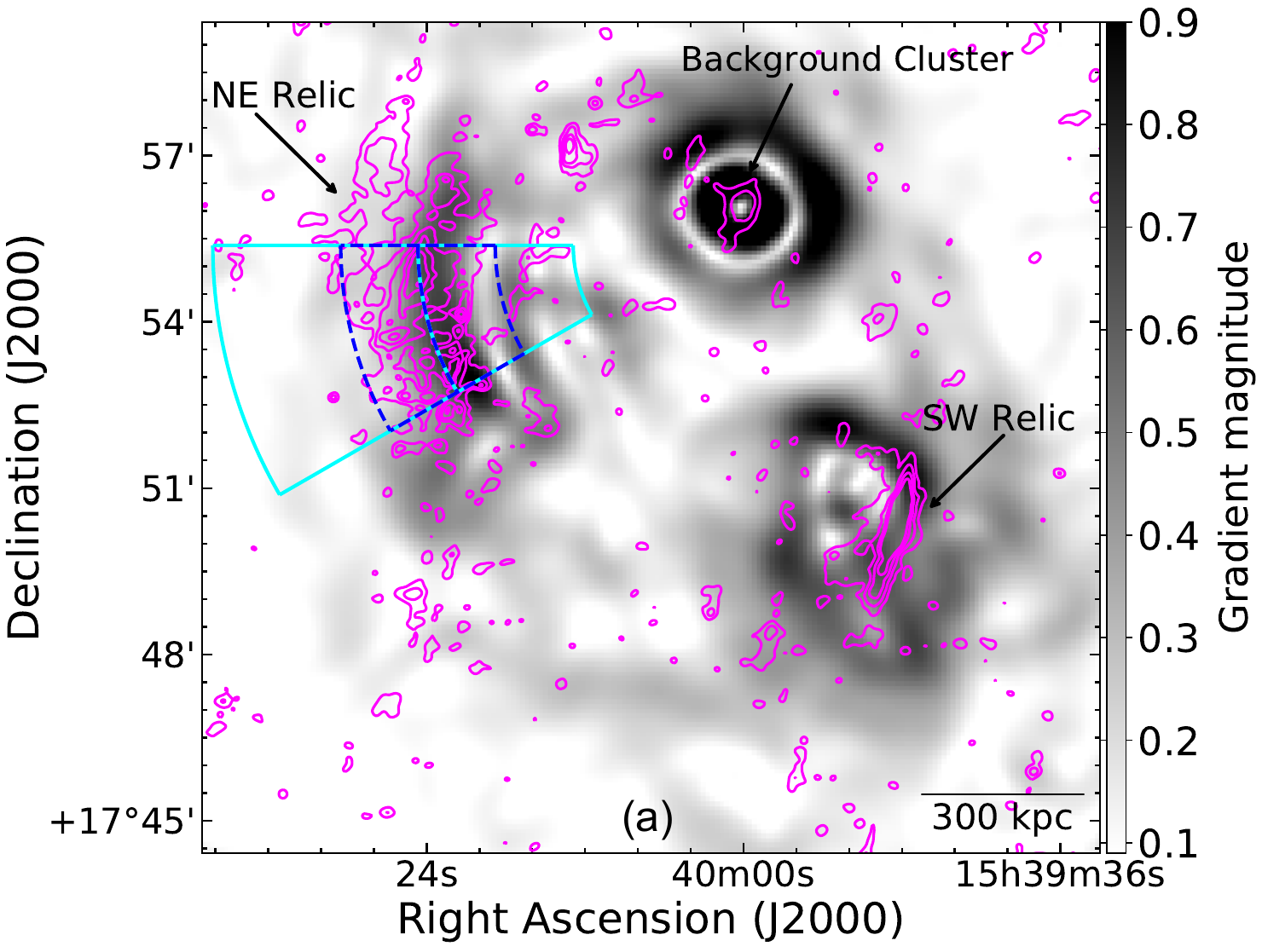}
\includegraphics[width=0.89\columnwidth, height= 23em]{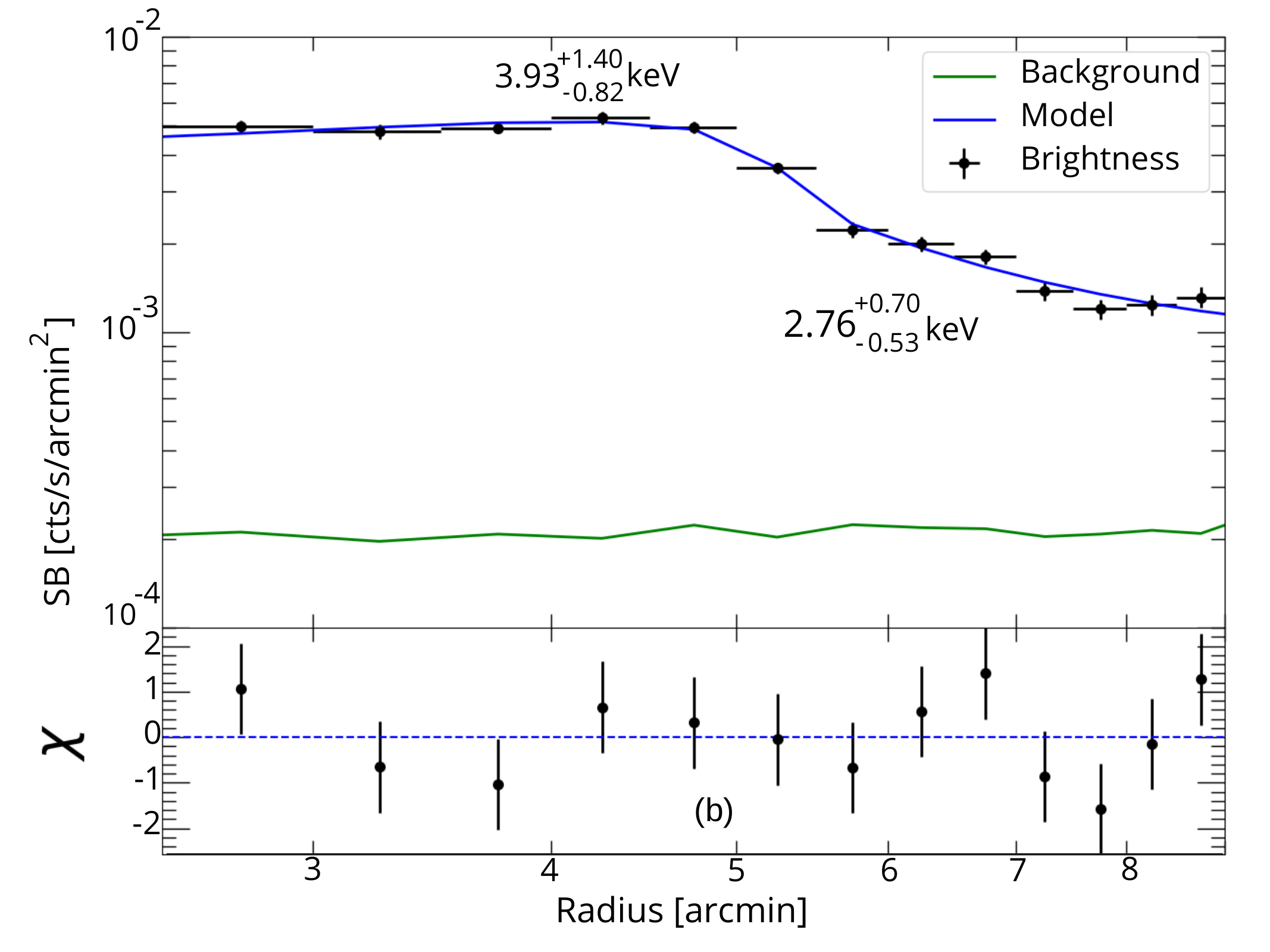} 
\begin{center}

\caption{(a): Background subtracted GGM filtered XMM-Newton X-ray image of A2108 overlaid with uGMRT point source subtracted radio contours in magenta. The cyan and blue wedge regions show the area used to quantify the SB discontinuity and temperature discontinuity, respectively, and the central arc represents the jump location. (b): The SB profile over the wedge region, as shown in Fig.~\ref{fig: A2108_shock_profile}a and the respective temperatures over the regions.}
\label{fig: A2108_shock_profile}
\end{center}
\end{figure*}

We observed the cluster with the GMRT during August 2021 (project code $40\_088$) as a part of uGMRT LOw-MAss Galaxy Cluster Survey (GLOMACS) using band-3 with 195 min on-source time. 3C 286 was observed as flux and bandpass calibrator and 1419+064 as phase calibrator. The cluster was observed in both GMRT Software Backend (GSB) and GMRT Wideband Backend (GWB) modes with 16 sec and 2.6 sec integration time and 32 MHz and 200 MHz bandwidth, respectively. 
Both the GSB and GWB data were analyzed using the Source Peeling and Atmospheric Modelling (\texttt{SPAM}; \citealt{refId0, Intema_refId0}) pipeline. At first, the GSB data was pre-calibrated using \texttt{SPAM}. During the pre-calibration process, the flux density scale was set using \citealt{scaife_10.1111/j.1745-3933.2012.01251.x} model. Thereafter, bad data editing, several loops of self-calibration, and finally, direction-dependent calibration are performed. The output primary beam-corrected widefield image was used as a sky model for our GWB data reduction.
The 200 MHz bandwidth of GWB data was split into 6 smaller frequency chunks, each of 33 MHz. These sub-bands were calibrated 
and processed individually with \texttt{SPAM} using the reference sky model extracted from the GSB image. Multi-scale multi-frequency synthesis (MFS) map was made using \texttt{WSClean v3.1.0} imager \citep{Offringa_2017MNRAS} in a joined-channel deconvolution mode using the calibrated visibilities from these 6 sub-bands.

We observed a faint diffuse emission signature near the Northeast of the cluster. 
To recover the whole structure and to achieve better image sensitivity, we used the available band-3 archival data of A2108 (Obs ID: 37\_018) with 132 min on-source time. The archival GWB data was split into 6 sub-bands (33 MHz), and the pre-calibration was done using SPAM. Each of the sub-bands was then merged with the pre-calibrated sub-bands from our GLOMACS observation data. The merged sub-bands were then further processed with SPAM. The full-resolution image of the cluster created with the combined datasets using WSCLean with Brigg's robust 0.5 in Fig \ref{fig: A2108_radio_xray}a (blue contours) shows both the faint diffuse emission as well as the compact sources present in the field. To remove the compact sources, we made high-resolution images for each of the merged sub-bands with Brigg's robust -1 applying an inner uv-cut of 3~k$\lambda$ (corresponding to $\sim 120$ kpc at the cluster redshift) using CASA task \textit{tclean}. The model visibilities from the high-resolution images were then subtracted from each of the sub-band visibilities.
We made the point source subtracted final radio map of the cluster with Brigg's robust 0.5, choosing the common uv-range of $<$20~k$\lambda$ for all sub-bands and smoothing it with a Gaussian taper of 7$\arcsec$ in \texttt{WSClean} (Fig. \ref{fig: A2108_radio_xray} b). 

\section{X-ray Observation} \label{X-ray}

We used archival XMM-Newton telescope observations (Obs ID: 	
0821810401) for this work. The data analysis was performed using the XMM-Newton Extended Source Analysis Software (\texttt{XMM-ESAS}), available in the SAS package. 
Further, we used \texttt{Xspec~v.12.0} for X-ray spectral analysis.
We performed a standard multi-step process to obtain clean EPIC data, spectral, and response files following the ESAS cookbook \footnote{XMM-ESAS cookbook : \url{https://heasarc.gsfc.nasa.gov/docs/xmm/esas/cookbook/}}.
Initially, the \textit{cifbuild} task was utilized to construct the calibration index file, followed by the ingestion of the observational data using the \textit{odfingest} task.
Subsequently, the \textit{epchain} and \textit{emchain} tasks were applied to perform the necessary calibration and filtering on the EPIC-pn and EPIC-MOS data, respectively. We used full observation sets (PN, MOS1, and MOS2) except CCDs 3 and 6 of MOS1 were excluded as they suffered from damage from hits by micro-meteorites.
Additionally, the pn-filter and mos-filter tasks were employed to apply specific filtering criteria for both instruments.
We have used an off-source region, which is far away from the centre of the cluster, as background.
Spectra were extracted using \textit{evselect} task, and the background was scaled using \textit{backscale} task.
All the spectra from PN, MOS1, and MOS2 were fitted simultaneously using XSPEC v12.0. The best-fitted temperature from the model Gaussian (0.47 keV) + Gaussian (1.75 keV) + constant*phabs*apec was noted for each region. The redshift, abundance (0.5 $Z_\odot$), and column density ($n_H$) were kept frozen during fitting. 

\begin{figure*}
\includegraphics[width=\textwidth]{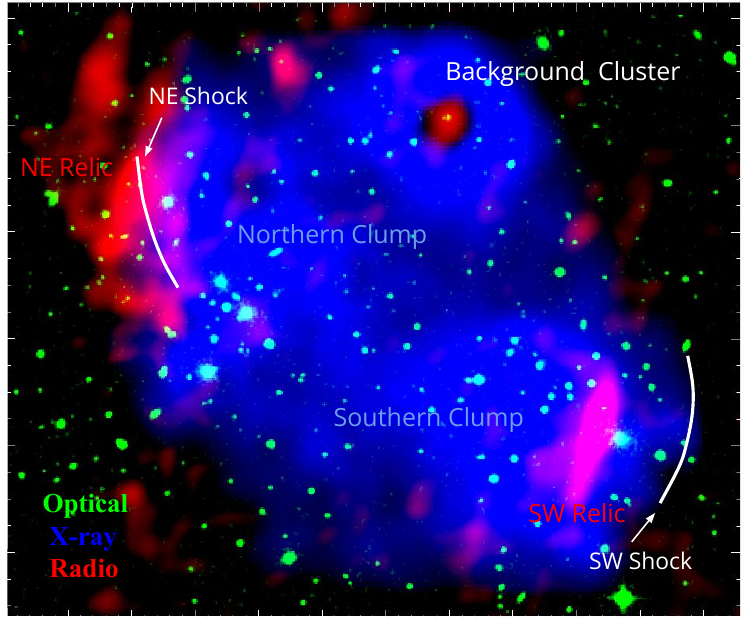}
\caption{Colour composite image of A2108 showing Pan-STARRS `$r$' band optical image in green, XMM-Newton X-ray in blue, and uGMRT point source removed radio image in red.}
\label{fig: A2108_rgb}
\end{figure*}

\section{Results from Radio Observation}
Fig.~\ref{fig: A2108_radio_xray}a shows the GMRT band-3 full resolution radio contours of the cluster A2108 obtained with Briggs robust 0.5 (\citealt{Briggs_1995AAS...18711202B}) and restoring beam $9.86\arcsec \times8.10\arcsec $ PA 87.28$\degree$ overlaid on Pan-STARRS `$r$' band optical image.
Two distinct diffuse emission features were observed in the cluster's southwestern (SW) and northeastern (NE) periphery that do not have any obvious optical counterparts (see Fig.~\ref{fig: A2108_radio_xray}a). The brightest regions of SW and NE emissions are located at a distance of 550~kpc and 438~kpc away from the cluster X-ray centre, respectively. The SW diffuse structure was previously identified by \citep{Schellenberger_2022ApJ} as a 200~kpc long radio relic. However, there was no mention of the NE diffuse emission in previous studies.
The location and morphology of the emission are indicative of a radio relic. 

From our observation, we detect the SW relic with the largest linear size (LLS) of $\sim300$~kpc and width of $\sim120$~kpc. The diffuse emission at NE shows a patchy structure and spans over a region of 610~kpc $\times$ 310~kpc. We estimated the flux density of the relics by measuring the total flux enclosed within 3$\sigma$ contour (where $\sigma=42~\mu$Jy/beam; Fig.~\ref{fig: A2108_radio_xray}b). We obtain the integrated flux density of $10.6\pm0.8$~mJy for the SW relic and $24.8\pm1.8$~mJy for the relic in the NE. Here, we note that the residual artefacts from the bright radio source D complicate the flux density measurement for the NE relic. Therefore, we also measured the flux density from the artefact-free brighter region falling under 6$\sigma$ contours for the NE relic and found a flux density of  $7.6\pm0.6$~mJy.
The error in the flux density was estimated following \citet{vanweeren_2021A&A}, where the map noise, uncertainty in absolute flux density calibration \citep[we used 7\% following][]{chandra_2004ApJ},  and uncertainty due to point source subtraction were taken into account. We have extrapolated the flux density of the relics to 1.4 GHz assuming a spectral index of $-1.3$ (typical for radio relics), and obtained a luminosity of  $P_{1.4 \mathrm{\ GHz|NE}} = 1.05 \pm 0.07\times10^{23}$~W~Hz$^{-1}$ and $P_{1.4 \mathrm{\ GHz|SW}} = 0.45 \pm 0.03\times10^{23}$~W~Hz$^{-1}$. We see the surface brightness (P$_{1.4\mathrm{GHz}}/\mathrm{LLS}^2$) of the NE relic is lower (5.55 $\times 10^{17}\mathrm{W~Hz}^{-1}\mathrm{Mpc}^{-2}$) compared to the SW relic (12.5 $\times 10^{17}\mathrm{W~Hz}^{-1}\mathrm{Mpc}^{-2}$). This can be one of the possible reasons for the non-detection of the relic in the previous observations, and the better UV coverage and sensitivity achieved with our analysis aided in the detection.
\section{Results from X-ray Observation}

We created an X-ray map of the cluster (shown in blue in Fig.~\ref{fig: A2108_rgb}) using the XMM-Newton X-ray observation, combining data from PN, MOS1, and MOS2 in the energy range of 0.4-10.0 keV. The X-ray map reveals the presence of multiple substructures within it. As previously noted by \citet{Schellenberger_2022ApJ}, the northwestern clump in the surface brightness (SB) map of the cluster is attributed to a background cluster. 

\subsection{Edge detection}

To locate any surface brightness discontinuity near the NE radio emission, we applied a Gaussian Gradient Magnitude (GGM) filtering method 
\citep{Sanders_2016MNRAS.460.1898S}.
This technique estimates the gradient magnitude or the rate of change of intensity values in an image using a Gaussian filter. The Gaussian filter smooths the input image, which helps in emphasizing the large-scale structures and reducing noise. The gradient magnitude is then calculated using derivative operators in the horizontal and vertical directions. The gradient magnitudes are then thresholded, where only pixels with intensity above a certain threshold value are retained. This further helps emphasize edges or boundaries with higher intensity while suppressing lower-intensity regions.

Fig.~\ref{fig: A2108_shock_profile}a represents the GGM-filtered image of A2108 overlaid with the radio contours. A sharp edge is seen at the northwest of the map due to the subtracted background cluster. 
Near the northeast, an arch-shaped SB edge can be observed coinciding with the diffuse radio emission. This sharp edge indicates discontinuities in the surface brightness, which can be caused by shock waves in the intracluster medium. 

\subsection{Surface Brightness and Temperature Discontinuity}

Further, we looked for any discontinuities associated with SB and temperature near the edge location. We extracted the SB profiles from multiple annuli spread over the wedge region near the NE relic and fit the SB profiles with the broken power law using PROFFIT \citep{Eckert_2011A&A...526A..79E}. We noticed an SB discontinuity coinciding with the edge detected with GGM filtering (Fig.~\ref{fig: A2108_shock_profile}b). Moreover, a tentative jump in temperature from 3.93$^{+1.40}_{-0.82}$ keV to 2.76$^{+0.70}_{-0.53}$ keV was also noticed near the edge. The wedge and the jump locations have been highlighted as cyan arcs for the SB jump and blue arcs for the temperature jump in Fig.~\ref{fig: A2108_shock_profile}a. 

Following \cite{Rahaman_2022MNRAS.509.5821R}, the shock Mach numbers were calculated using the Rankine-Hugoniot shock jump conditions \citep{Rankine_1870RSPT, hugoniot1887memoire, hugoniot1889propagation}. The SB jump was calculated following the equation,
\begin{equation}\label{eq1}
    \frac{\rho_2}{\rho_1} = C = \frac{(1+\gamma)\times \mathcal{M}_\mathrm{SB}^2}{2+(\gamma - 1)\times \mathcal{M}_\mathrm{SB}^2}
\end{equation}
where $\rho_1$ \& $\rho_2$ are densities at pre and post-shock regions respectively.  Considering $\gamma = 5/3$ for monoatomic gas, we get a shock Mach number from SB jump across the shock edge, $\mathcal{M}_\mathrm{SB} = 1.42^{+0.14}_{-0.14}$.

The Mach number from temperature jump was calculated following the equation 
\begin{equation} \label{eqn2}
    \frac{T_2}{T_1} = \frac{(5\mathcal{M}_\mathrm{T}^2 - 1)(\mathcal{M}_\mathrm{T}^2 + 3)}{16\mathcal{M}_\mathrm{T}^2}
\end{equation}
where $T_1$ and $T_2$ are the pre-shock (upstream) and post-shock (downstream) temperatures, respectively. 
The Mach number from temperature jump was found to be $\mathcal{M}_\mathrm{T} = 1.43^{+0.12}_{-0.02}$.


\section{Discussion}

Simulations show peripheral radio relics are to be generated at diametrically opposite sides due to cluster merger occurring in the plane of the sky \citep{vanWeeren_2011MNRAS, Bruggen_2012MNRAS}. However, there are only a few double relic systems reported so far, with most of the detection lying in clusters with M$_{500}\geq 5 \times 10^{14}\mathrm{M}_{\odot}$
\citep{Bonafede_2017MNRAS}. So far, $\sim$10 clusters have been detected with double relics in the low mass regime \citep{Bonafede_2017MNRAS, Jones_2023arXiv, Koribalski_2023arXiv}. In addition to the previously detected relic at the southwest of A2108 \citep{Schellenberger_2022ApJ}, the detection of another diffuse structure at the northeastern periphery of the cluster reveals this cluster as a potential double relic system originated from the cluster merger-driven shocks. The X-ray surface brightness map of A2108 shows an elongation along northeast (NE) and southwest (SW), suggesting the potential merger axis. From the reported radius ${r}_{200} =1.6$~Mpc \citep{Jensen_2012MNRAS}, following \citet{Wen&han_2013MNRAS}, we estimated the $r_{500}$ to be $\sim$1 Mpc for A2108 cluster. The relics of A2108 are also situated at a distance of $\sim 1$~Mpc in the sky plane and have a significantly different size (300~kpc $\times$ 120~kpc for the SW relic and 610~kpc $\times$ 310~kpc for the NE relic). Correlation studies to understand the connection between host cluster mass and radio relic luminosity at 1.4 GHz have been performed previously \citep{gasperin_2014MNRAS, Kale_2017MNRAS}. In a recent work, \citet{Duchesne_2021PASA} updated the scaling relation by incorporating double relics only. We revisited the correlation of radio relic power at 1.4 GHz with the host cluster mass for both single and double relics. Our sample comprises relics with known 1.4 GHz radio power from \citet{gasperin_2014MNRAS, Kale_2017MNRAS, HyeongHan_2020ApJ, Locatelli_2020MNRAS.496L..48L, Paul_2021MNRAS, Duchesne_2021PASA, Lee_2022ApJ, chatterjee_2022AJ}. We used the Bivariate Estimator for Correlated Errors and Intrinsic Scatter (BCES) method to perform a linear regression on $P_{1.4~\mathrm{GHz}}$ and $M_{500}$. BCES accounts for errors in both the dependent and independent variables and for the intrinsic scatter of the data and thus is a robust approach. The scaling relation, $\mathrm{log(P_{1.4~GHz}) = A log (M_{500}) + B}$, yielded best-fit parameters A = 3.07 $\pm$ 0.26 and B = -20.86 $\pm$ 3.88, consistent with estimates incorporating the clusters with double relics \citep{Duchesne_2021PASA}. We see that the NE relic is $\sim 3$ times under-luminous, and the SW relic is $\sim 9$ times under-luminous than what is expected from the relic mass-luminosity correlation.

\begin{figure}
\includegraphics[width=0.99\columnwidth]{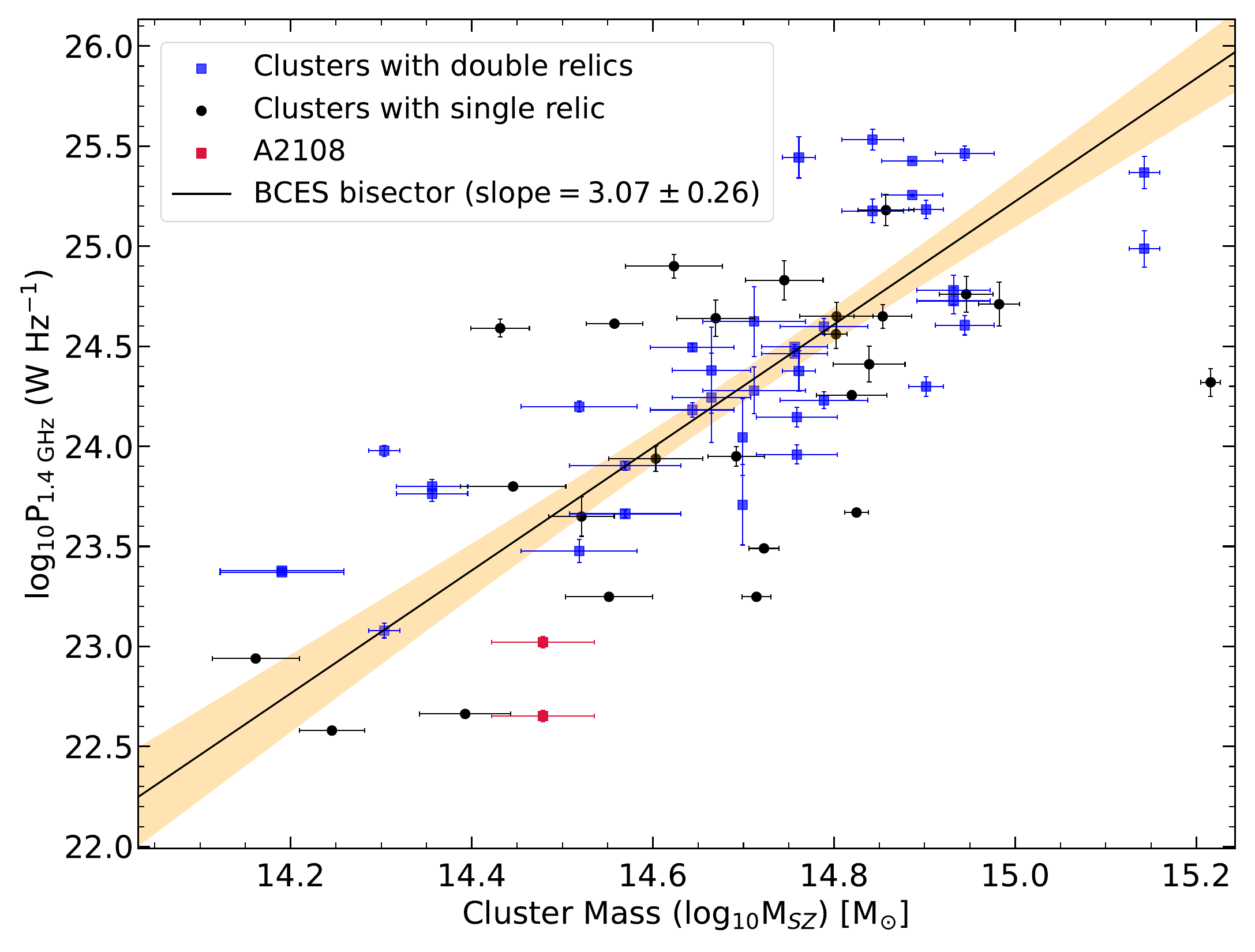}
\caption{The 1.4 GHz power of radio relics available from literature \citep{Kale_2017MNRAS,HyeongHan_2020ApJ,Locatelli_2020MNRAS.496L..48L,Paul_2021MNRAS, Duchesne_2021PASA, Lee_2022ApJ, chatterjee_2022AJ} is plotted against the host cluster mass. The blue squares show the double relics, and the black circles show the single relic systems. The red squares show the relics of A2108.}
\label{fig: A2108_pm}
\end{figure}
\begin{figure}
\includegraphics[width=0.99\columnwidth]{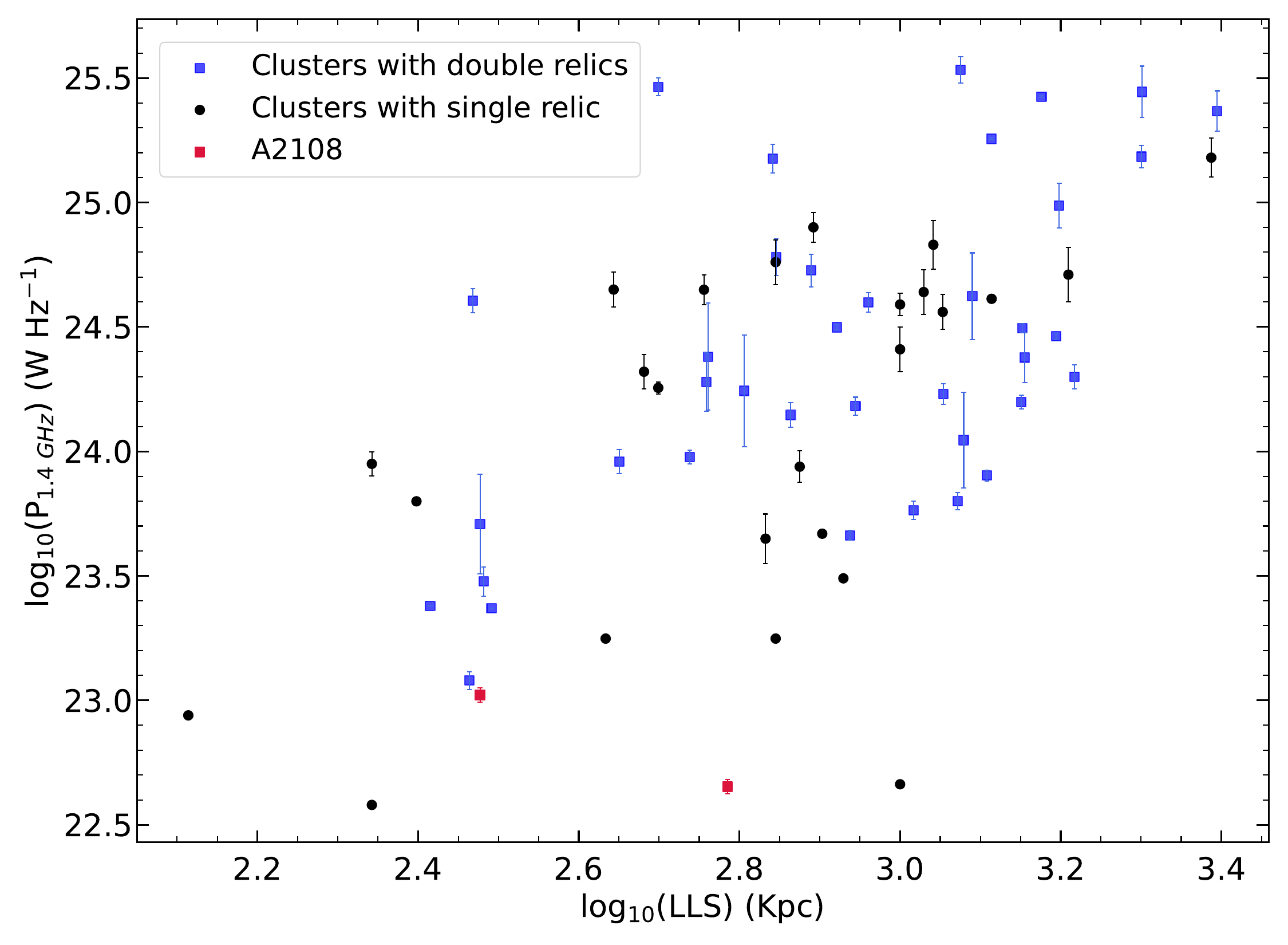}
\caption{The 1.4 GHz power of radio relics available from literature \citep{Kale_2017MNRAS,HyeongHan_2020ApJ,Locatelli_2020MNRAS.496L..48L,Paul_2021MNRAS, Duchesne_2021PASA,Lee_2022ApJ, chatterjee_2022AJ} is plotted against the relic LLS. The blue squares show the double relics, and the black circles show the single relic systems. The red squares show the relics of A2108.}
\label{fig: A2108_p_lls}
\end{figure}

Additionally, we compared the radio power and the LLS of the relics with the other relics. In Figure~\ref{fig: A2108_p_lls}, we see that both the relics, especially the SW relic, are quite under-luminous compared to most of the double relic systems. The LLS of double relics generated in binary cluster merger roughly scales with the sub-cluster masses as shown by \citet{vanWeeren_2011MNRAS}. However, it is important to consider the influence of the local environment in the relic formation and potential observational biases introduced due to projection effects \citep{vanweeren_2011A&A, Jones_2021MNRAS, Lee_2022ApJ}. Despite the fact that the sub-clusters of A2108 have comparable masses \citep{Schellenberger_2022ApJ}, the LLS and breadth of the NE relic are significantly different (larger by a factor of two and three, respectively) compared to the SW relic. Moreover, the surface brightness ratio of the two relics is also not close to unity, as seen for most of the double relic systems \cite{Lee_2022ApJ}. These factors indicate the relics may not have the same origin. While the origin of the SW relic could be justified by the passage of shock front \citep{Schellenberger_2022ApJ}, the morphology of the NE relic is suggestive of the influence of other secondary factors in the origin of the relic.

The X-ray SB map of the cluster also reveals interesting features in the ICM of A2108. Two X-ray bright sub-clusters at NE and SW and a faint X-ray emitting region between them, possibly from the trailing material of the sub-clusters, can be observed in Fig.~\ref{fig: A2108_rgb}. 
The low radio luminosity of the NE relic could be explained by the presence of a weak shock ($\mathcal{M}_\mathrm{SB} = 1.42^{+0.14}_{-0.14}$) detected near the relic from X-ray analysis. 
Interestingly, this shock and the previously detected shock 200 kpc away from the SW relic \citep{Schellenberger_2022ApJ} appear in non-diametrically opposite directions (Figure ~\ref{fig: A2108_rgb}), indicating a complex merger geometry. The cluster X-ray peak is sensitive to density, whereas the SZ peak is sensitive to pressure along the line of sight. Therefore, an offset is expected for merging clusters \citep{Molnar_2012ApJ, Zhang_2014ApJ}. However, the offset of the Peaks observed in A2108 (Fig~\ref{fig: A2108_radio_xray}a ) is roughly perpendicular to the merging axis, which is unusual given the cluster merging axis in the NE SW direction. The misalignment observed in optical galaxy density and the X-ray clumps also hint towards a non-linear merger axis \citep{Schellenberger_2022ApJ}. We hypothesize that this feature can be observed due to an off-axis merger. Previously, off-axis merging has been predicted for galaxy clusters for instance, A115 \citep{Hallman_2018ApJ, Lee_2020ApJ} and A141 \citep{Caglar_2018MNRAS}. Interestingly, both A115 and A141 both host diffuse radio emission. Whereas A115 was reported to host a single radio relic \citep{Govoni_2001A&A} with also a shock detected at the relic location \citep{Botteon_2016MNRAS.460L..84B},  A141 was reported to host a radio halo (see e.g. \citealt{Duchesne_2021PASA_b}).

The brighter emission of the NE relic is also slightly misaligned with the shock front. Moreover, the NE relic is more expanded in the shock downstream region, which is peculiar to be formed by the DSA mechanism and considering the low Mach number detected via X-ray observation. 
There are multiple radio galaxies near the NE relic (highlighted with red `+' in Fig.~\ref{fig: A2108_radio_xray}b). However, due to the lack of spectroscopic redshift information for the host galaxies, cluster membership cannot be ascertained. Apart from the BCGs, we find the radio galaxy E with a possible optical host (SDSS J154017.98+1753; \textit{z} = 0.09104) situated within 27 kpc distance at sky plane. No optical host was found for galaxy D. The re-energisation of the old electron population from these galaxies possibly has played a crucial role in the generation of the NE relic. The stronger evidence in favour of this scenario comes from the presence of patchy morphology of the relic. However, a detailed radio spectral index study would be required to shed more light on the particle acceleration phenomena near the relics.

\section{Conclusion}

In this paper, we report the discovery of a new relic in the northeast of low-mass galaxy cluster A2108 and confirm the presence of the other relic in the southwest with higher significance using uGMRT band-3. This new discovery places A2108 among the few low-mass clusters with double radio relics. The radio power of both the relics extrapolated to 1.4 GHz was found to be considerably low, making them outliers in the mass-luminosity plane. However, the non-axial X-ray shock positions, the significant size difference of NE (610 kpc $\times$ 310 kpc) and SW (300 kpc $\times$ 120 kpc) relics, their distinct morphologies, etc., helped us argue in favour of contrasting formation mechanism for these two relics. 
 
Moreover, using the XMM-Newton science archive observation, we detect the presence of surface brightness jump and tentative temperature jump in the midst of the NE relic, indicating a weak shock ($\mathcal{M}_\mathrm{SB} = 1.42^{+0.14}_{-0.14}$, $\mathcal{M}_\mathrm{T} = 1.43^{+0.12}_{-0.02}$) usually disfavouring DSA origin for the relic. While the origin of the SW relic can be well justified from the passage of shock, the presence of multiple radio galaxies and the expanded structure of the NE relic suggest that the fossil electrons from radio galaxies play an important role in the generation of this source. Nevertheless, more clarity in understanding the underlying physical process of formation of these relics would need the information on the spectrum, requiring multi-frequency observations.

\section*{Acknowledgements}

We thank IIT Indore for giving out the opportunity to carry out the research project. MR acknowledges support from the National Science and Technology Council of Taiwan (MOST 109-2112-M-007-037-MY3; NSTC 112-2628-M-007-003-MY3). This  research is supported by DST-SERB through ECR/2017/001296 
grant awarded to AD. RK acknowledges the support of the Department of Atomic Energy, Government of India, under project no. 12-R\&D-TFR-5.02-0700 and from the SERB Women Excellence Award WEA/2021/000008. SP wants to thank the DST INSPIRE Faculty Scheme (code: IF-12/PH-44), during which the GLOMACS project was initiated. We thank the staff of the GMRT who have made these observations possible. The GMRT is run by the National Centre for Radio Astrophysics of the Tata Institute of Fundamental Research. This research made use of Astropy,\footnote{http://www.astropy.org} a community-developed core Python package for Astronomy \citep{Astropy_2013A&A,astropy_2018}, Matplotlib \citep{Matplotlib_Hunter:2007}, and APLpy, an open-source plotting package for Python \citep{APLpy_2012ascl.soft08017R}.

\section*{Data Availability}
The data used for this work is available in the GMRT Online Archive (\url{https://naps.ncra.tifr.res.in/goa/data/search} and XMM-Newton Science Archive (\url{http://nxsa.esac.esa.int/nxsa-web/#home}).



\bibliographystyle{mnras}
\bibliography{example} 








\bsp	
\label{lastpage}
\end{document}